\documentclass[%
 aip,
 floatfix,
 amsmath,amssymb,
 reprint,%
]{revtex4-1}

\usepackage{graphicx}
\usepackage{amsfonts, amsmath, amstext, amssymb, amsfonts, amsxtra}
\usepackage{braket}
\usepackage{bbm}
\usepackage{bbold}
\usepackage{dcolumn}
\usepackage[protrusion=true,expansion=true]{microtype}
\usepackage{wasysym}
\usepackage{cleveref}
\usepackage{color}
\usepackage[normalem]{ulem}
\usepackage{bm}
\usepackage[utf8]{inputenc}
\usepackage[T1]{fontenc}
\usepackage{mathptmx}
\usepackage{float}
\usepackage{array}
\usepackage{booktabs}
\usepackage{url}

\newcolumntype{C}[1]{>{\centering\arraybackslash}p{#1}}

\begin{document}

\title{Game-driven random walks: Survival time statistics}

\author{M.I.~Krivonosov}
\email{krivonosov@itmm.unn.ru}
\affiliation
{ 
    Department of Applied Mathematics, Lobachevsky University, 603022, Nizhny Novgorod, Russia
}
\affiliation
{ 
    Mathematical Center, Lobachevsky University, 603022 Nizhny Novgorod, Russia
}

\author{S.N.~Tikhomirov}
\affiliation
{ 
Department of Applied Mathematics, Lobachevsky University, 603022, Nizhny Novgorod, Russia
}

\author{S.~Denisov}
\email{sergiyde@oslomet.no}
\affiliation
{ 
    Department of Computer Science, Oslo Metropolitan University, N-0130, Oslo, Norway
}
\affiliation
{ 
NordSTAR – Nordic Center for Sustainable and Trustworthy AI Research, Oslo N-0166, Norway
}
\affiliation
{ 
    Mathematical Center, Lobachevsky University, 603022 Nizhny Novgorod, Russia
}

\date{\today}

\begin{abstract}
Random walks are powerful tools to analyze  spatial-temporal patterns produced by living organisms ranging from cells to humans. At the same time, it is evident that these patterns are  not completely random but are results of a convolution of organisms' sensor-based information processing
and motility. The complexity of the first  component is reflected in the statistical characteristics of trajectories produced by an organism  -- when it is, e.g., foraging or searching for a mate (or a pathogen) -- and therefore some knowledge about the component can be obtained by analyzing the  trajectories with the standard toolbox of methods used for random walks. 
Here we consider trajectories which appear as the results of a game played by two players on a finite square lattice.  One player wants to survive, i. e., to stay within the interior of the square,  as long as possible while another one wants to reach the  adsorbing boundary. A game starts from the center of the square and every next movement of the point is determined by independent strategy choices made by the players. The value of the game is the survival time that is the number of steps before the adsorption happens. We present the results of a series of experiments  involving both human players and an autonomous agent (bot)  and concentrate on the probability distribution of
the survival time. This distribution indicate that the process we are dealing with is more complex than the standard random walk. 
\end{abstract}

\maketitle

\begin{quotation} There are close ties between 
random walks and game theory. A one-dimensional random walk on a finite discrete interval is famously interpreted as a story of two gamblers playing the same random game (f.e., by flipping a coin)  again and again until  one of the gambler’s capital is exhausted and he is ruined1 \cite{ruin1,ruin2}. In the random-walk framework, this is a situation when the random walker -- whose position is determined by the current capital 
of one of the players -- reaches one of the two boundary points and is adsorbed there. The ruin time is therefore first-passage or first-adsorption time, one of the key concepts of the random walk theory \cite{redner}. 

Assume now that the game is not completely random but the corresponding  transition probabilities are controlled by the players. This is the idea of "games of survival" proposed by Hausner \cite{survival1} and Peisakoff \cite{survival2} in their notes in 1952. Similar to the original gambler ruin's process \cite{ruin3D}, survival games can be generalized to two and three dimensions and finally formulated in the most general way as a game-driven motion over some finite domain with a border \cite{survival3}. In such a game one of the players tries to avoid hitting the border while  his opponent wants to reach the border as fast as possible. The game is specified with a matrix with a certain number of rows (strategies of the first player) and columns (strategies of the second player). On every step, the players  choose independently their strategies, that is the first player chooses a row and the second one - the column,  and the move of the point is determined by the corresponding element of the matrix. The game is over once the point reached the boundary. 

Even though the idea of such games was formulated  more then seventy years ago \cite{survival3}, no quantitative results were presented until now. This is not a surprise as such games are characterized by an immense complexity~\cite{complexity}.  
Here we present an experimental realization of a two-dimensional survival game. Our idea is to consider the game as a first-passage process~\cite{redner} and analyze the collected game histories from the corresponding perspective, by looking at the statistics of the first-passage or, more proper, the survival time.  
\end{quotation}

\section{Introduction}

From the very start, random walks were interpreted in biological and behavioral
terms, and now Pearson's drunkard~\cite{pearson}is no less famous than Schr{\"o}dinger's cat. Different type of random walks \cite{levy,int} are used as foundations to build optimal search theories and explain the spatial-temporal patterns  produced by foraging animals and people exploring Disney World  park\cite{park}. 

However, whenever a new random-walk model is introduced, a round of discussion
about model's validity and the importance of all details neglected during the model's formulation, immediately starts. E.g., it is often mentioned that  search landscapes are not uniform but patchy and often exhibit fractal-like patterns, which might induce a power-law scaling in foraging trajectories \cite{patchy1,patchy2}, or or that people usually follow streets or existing trails which form a very correlated structure\cite{trail}, etc. Another words, there are always factors and features that are not taken into account and thus there is a always a volume for a discussion on whether they are important and relevant, see.f.e., Refs.~\cite{pyke,denny}. Therefore it would be beneficial to \textit{design} a process, perform series of experiments, with organisms or humans, and collect the statistically sounding amount of data which then can be analyzed.

In case of humans, game theory provide a possibility to realize this program. The idea of game-driven or "game-type"~\cite{survival3} random walks, when players determine the move of a point by independently choosing strategies, looks like a good opportunity.

Here we present the results collected for a two-dimensional game-driven process,
by using a mobile app "Random Walk Game" \cite{app}. The app
provides two players with a possibility to play regardless of their location, make brakes and interruptions, and challenge each other for a round a game by inviting an opponent from the list of registered players.  The app allowed us to collect data more quickly and with less burden than traditional methods (such as lab sessions) allow. Additionally, the app provides  a human player with a possibility
to play anytime against a bot, who is simply flipping a fair coin on every round.

We analyzed the collected game histories from a point of view of survival time statistics, by comparing the survival time histograms with the distributions corresponding to random walks and some basic strategies. The experimental  histograms bear several features which demonstrate that the game-driven walks 
....

\section{Game}

The Random Walk Game is a game played on a square lattice, with vertices $(i,j)$, $i,j \in {0,1,...,N}$, where $N$ is an even number,  by two antagonistic players, $A$ and $B$. The game always starts from the center of the square, i.e., from the vertex $(\frac{N}{2},\frac{N}{2})$.
Both players can  see the current position of a point (marker) on the field.  The game field is a square lattice  with an odd length $n$ that consists of internal and adsorbing (boundary) lattice sites. Initially the point is placed at the center of the square. 

\begin{figure}[b]
    \begin{center}
    \includegraphics[width=0.7\columnwidth,keepaspectratio,clip]{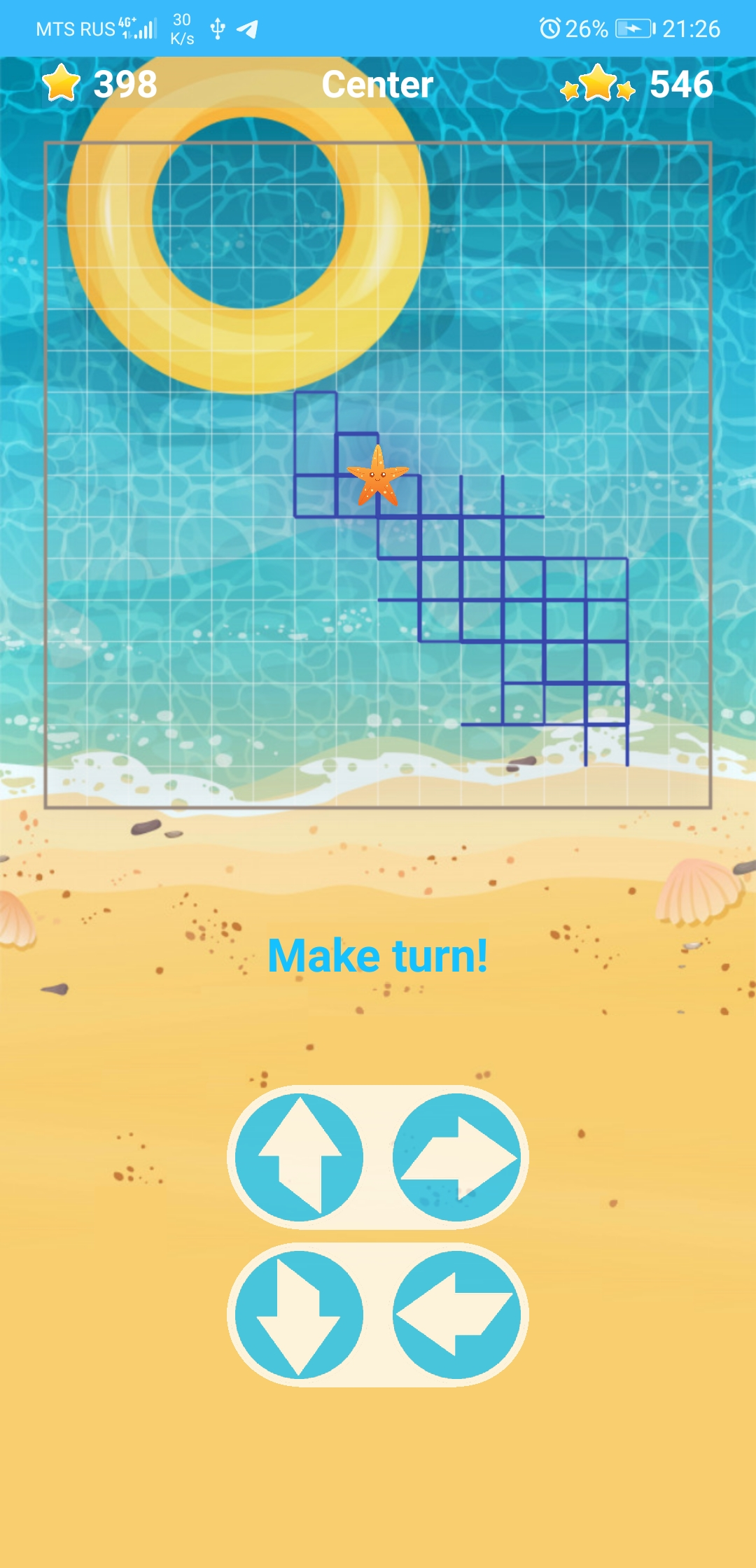}
    \caption{
        (Color online) Random Walk Game app screenshot. The header line consists of the current number of turns, their goals and the number of turns in the longest game. Player see the game field, the piece position and the trajectory of piece. At the bottom of the screen shown the control matrix 2 by 2 that defines the outcome of joint strategy choices of the player A and B. The rows is a possible choice of strategies for the player A, and columns -- for the player B. The resulting movement direction is determined by the arrow in the cell located in the corresponding row and column.
    }  
    \label{fig:1}
    \end{center}
\end{figure}

During the each step of the game, players control the movement of a point by making independent choice one of two possible strategies. Information about the possible movement (determined by the joint choice of the players) are represented by a matrix which is known by both players (see Fig.~\ref{fig:1}). The aim of the first player (A) is to keep the point within inside the square, i.e., at one of the internal sites, as long as possible, while the aim of the second player (B) is to reach the absorbing boundary, i.e., one of the adsorbing sites, as quickly as possible. 

\section{Experiment}

\subsection{Mobile app}
To organise the experiment part we developed a mobile application available at Google Play Store (https://play.google.com/store/apps/details?id= com.scigames.RWGame) and at AppStore (https://apps.apple.com/us/app/random-walk/id1564589250). The app provides two game modes: playing against another player (PvP) or playing against the environment (PvE). In both modes, app transmits player choices over the Internet to the web-server and receives resulting piece movements. All player choices and piece movements were collected in the server database for further analysis.
Choices of the environment in the PvE mode are determined by the sequence of pseudo-random numbers computed by Mersenne Twister random number generator (implemented in PHP 7.4 as mt\_rand function). The game implementation is only provides the game field size $17 \times 17$.

\subsection{Participants}
Participants of the reported study were aged from 16 to 52 years. Young players were volunteers recruited from the student population of Lobachevsky State University and Higher School of Economics. Older adults were recruited from prof. S. Denisov colleagues in different universities across the world.

\subsection{Experiment Design}
Three design setups were used to organise PvP game sessions: 
\begin{enumerate}
    \item Students sat in the same room and played without communications for 1 hour. The participant interest were the obtaining of the highest score in the room. Player pairs are selected based on similarity of their skills in the competitive programming.
    \item Students sat in the same room and played without communications for 2 hours. The participant were arranged by the sum of weighted scores of played games. The longer game - the higher weight associated with this game for the center player (A) and opposite for the border player (B).
    \item Participants sat at home and played without communications for around 30 minutes per day. The participant driver were their interest in the highest (lowest) score in the particular game.
\end{enumerate}

To organise PvE game sessions we make two setups:
\begin{enumerate}
\item Participants played on their own against environment and can pause the game at any moment. The session were continued for a one month. Players were ranked by the ratio of exponentially moving average (EMA) score of center goal games to the EMA score of border goal games. Player ranking table were available online to participants. 
\item Participants played on their own against environment and can pause the game at any moment. The same ranking table as defined in the previous setup was used. The task of participants is to beat the highest score.
\end{enumerate}

\section{Theoretical Analysis}

In this section we considered analysis of the pure random walk in case of two players picking randomly their strategy (BvB), the case of playing against the environment (PvE) and playing against another player (PvP). The game were investigated in the Markov Chain framework that describes memoryless stochastic process \cite{markov_chain}. The state of the system is a coordinate vector $w_k = (x_k, y_k)$ at the $k$-th turn, where position corresponds to the node of square lattice and coordinates vary in the ranges: $-\lfloor n/2 \rfloor \leq x_k \leq \lfloor n/2 \rfloor$, $-\lfloor n/2 \rfloor \leq y_k \leq \lfloor n/2 \rfloor$. Overall there are $n^2-4$ reachable nodes, $r=4(n-2)$ of which are corresponds to absorbing states $(|x_k|=\lfloor n/2 \rfloor \lor |y_k|=\lfloor n/2 \rfloor)$ and the remaining $s=(n-2)^2$ to transient states $(|x_k|<\lfloor n/2 \rfloor \land |y_k|<\lfloor n/2 \rfloor)$, see Fig.~\ref{fig:1}. Although, there are 2 parity cases of the lattice size, we consider only odd sizes due to the symmetry about the center that corresponds to the starting state. The minimum possible size of the game field is $3 \times 3$. Let us arrange nodes according to the horizontal scanning. Then, theory introduced transition matrix \cite{gagniuc_2017} $\rm P$ with elements corresponding to probabilities of hops $P((i, j) \rightarrow (x, y))$ from the state $(i, j)$ to $(x, y)$.

\subsection{Absorbing Markov Chains}
The general Markov chain theory does not cover part of absorption processes \cite{1}. However, the absorbing Markov Chain extension theory is dealing with such a case and provides a way to compute the time statistics.
In accordance to this extended theory the transition matrix ${\rm P}$ can be represented as a block matrix:
\begin{equation}
    \begin{aligned}
    {\rm P}=
      \begin{bmatrix}
        Q & R \\
        {\bf 0} & I_r
      \end{bmatrix}
    \label{eq:P}
    \end{aligned}
\end{equation}
where $Q$ is a $s$-by-$s$ matrix corresponding to transitions between internal states, $R$ is a nonzero $s$-by-$r$ matrix corresponding to transitions from internal states to absorbing states, and last two matrices are corresponding to loops in the absorption states: ${\bf 0}$ is an $r$-by-$s$ zero matrix, and $I_r$ is the $r$-by-$r$ identity matrix \cite{2}.

Then, we compute fundamental matrix $N$:
\begin{equation}
    \begin{aligned}
    N=\sum_{k=0}^{\infty} Q^k=(I_s-Q)^{-1}
    \label{eq:N}
    \end{aligned}
\end{equation}
where $I_s$ is the $s$-by-$s$ identity matrix. The $(i, j)$ entry of matrix $N$ is the expected number of times the chain is found in state $j$, given that the chain started in state $i$.

Based on the property of fundamental matrix we obtain the expected number of steps before being absorbed starting in transient state $i$:
\begin{equation}
    \begin{aligned}
    T=N{\bf 1}
    \label{eq:T}
    \end{aligned}
\end{equation}
where ${\bf 1}$ is a column vector whose entries are all 1.

Since the starting position of the game is a center of square lattice the resulting expected number of steps before being absorbed ${\rm \bf t_n}$ corresponds to entry of $T$ for the $(0, 0)$ state, where $n$ is a size of field.

\subsection{Game in terms of Markov Chains}
Applying of Markov Chain theory requires the certain values of the transition probability between states. In general, relying on memoryless property of Markov Chain, we define mixed strategy $S_{ij}^p$ determined by Bernoulli distribution ${\bf \sigma}_{ij}^p$ over two pure strategies for the player $p \in \{A, B\}$ in the state $(i, j)$:
\begin{equation}
    \begin{aligned}
    {\bf \sigma}_{ij}^p(s)=
    \begin{cases}
        f_{ij}^p, &\mbox{if } s = 0,\\ 
        1-f_{ij}^p, &\mbox{if } s = 1,\\
    \end{cases}
    \label{eq:sigma}
    \end{aligned}
\end{equation}
where $f_{ij}^p \in [0, 1]$ is a probability of picking first strategy $(s=0)$.

This in turn allows us to determine the transient probability from the state $(i, j)$ to $(x, y)$ for the corresponding mixed strategies:
\begin{equation}
    \begin{aligned}
    P& \left( (i, j) \rightarrow (x, y) \right) = \\
    &=\begin{cases}
        0, &\mbox{if } |i-x|+|j-y| \neq 1,\\ 
        f_{ij}^A \left(1-f_{ij}^B\right), &\mbox{if } x=i \land y=j+1, {\bf \rightarrow}\\
        \left(1-f_{ij}^A\right) f_{ij}^B, &\mbox{if } x=i+1 \land y=j, {\bf \downarrow}\\
        \left(1-f_{ij}^A\right) \left(1-f_{ij}^B\right), &\mbox{if } x=i \land y=j-1, {\bf \leftarrow}\\
        f_{ij}^A f_{ij}^B, &\mbox{if } x=i-1 \land y=j, {\bf \uparrow}\\
    \end{cases}
    \label{eq:transition}
    \end{aligned}
\end{equation}

The connectivity in the system is defined by transitions to 4-neighbor states on the square lattice. In this way the first case of transient probability denotes the zero probability between unconnected states, and others are transition probabilities to neighbor states in the particular direction.

The randomly picking strategy case corresponds to equal probability of pure strategies of both players: $f_{ij}^A=1/2$, $f_{ij}^B=1/2$. Therefore, the transient probability is equal to $1/4$ for neighbour state transition and zero otherwise. Applying absorbing Markov Chain formalism we computed the mean absorption time ${\bf t_n^{BvB}}$ depending on the field size $n$.

The last 2 cases of playing against the environment (PvE) and playing against another player (PvP) explicitly determines only environment strategy which implies an equiprobable choice. Despite this, strategies of participants are unclear. Therefore, in order to assess behavior we assume memorylessness of participant choices and estimate relative frequencies of their strategies depends on the position in the square lattice. Next, we interpreted experimentally obtained relative frequencies as probabilities and computed the mean absorption time for both cases: ${\bf t_n^{PvE}}$ and ${\bf t_n^{PvP}}$ depending on the field size $n$.

\subsection{Probability evolution modelling}

Although, absorbing Markov Chains are let us find the mean absorption time of the random walk, the exact distribution of absorption time and spatial distribution cannot be obtained in this framework. To solve this problem we define a mathematical model based on a Markov chain property of the game. Let us define the probability $W_{ij}^{k}$ of detecting a piece in a state $(i, j)$ on the $k$-th turn. The $(0, 0)$ initial state of the game corresponds to the initial condition of finding piece in the $(0, 0)$ state with probability one $W_{00}^{0}=1$ and with probability zero in other states $W_{ij}^{0}=0$, where $(i, j) \neq (0, 0)$. The sequential applying of the modified transition matrix ${\rm \widetilde{P}}$ to the state probability vector ${\bf w^{k}}$ determines the process of the state distribution propagation:

\begin{equation}
    \begin{aligned}
    {\bf w^{k+1}}={\rm \widetilde{P}}{\bf w^{k}}, k \geq 0
    \label{eq:evolution}
    \end{aligned}
\end{equation}

At the each turn we computed exact probability $W_{ij}^{k}$ that piece finishes in the particular absorption state at this turn, where $(i, j) \in {\bf B}$ -- boundary states. To do this, we modified transition matrix ${\rm P}$ in such a way that the reaching the boundary state excludes the piece from the system. Then, the probability of piece being absorbed at the $k$-th turn can be computed using the following formula:

\begin{equation}
    \begin{aligned}
    p_{\rm abs}^{k}=\sum_{(i, j) \in {\bf B}} W_{ij}^{k},
    \label{eq:timedistr}
    \end{aligned}
\end{equation}
where {\bf B} is a set of boundary states.

The modelling of probability evolution also provides information about the spatially ending distribution. Computing of probabilities for absorption states were performed by the approximating of the corresponding series:
\begin{equation}
    \begin{aligned}
    p_{ij}^{\rm abs}=\sum_{k=0}^{\infty} W_{ij}^{k}
    \label{eq:spacedistr}
    \end{aligned}
\end{equation}
where $(i, j) \in {\bf B}$. Due to the exponential decrease of absorption probability with increasing the number of turns, summation can be stopped once the given precision is reached.

Another interesting property of the game is an influence of the turn parity on the probabilities. In view of the fact that the square lattice is a bipartite graph, the process of propagation at each turn jumps from the one part of the graph to another. Depends on the parity of the turn number we denote the even and odd part of the graph. The corresponding states are arranged on the lattice in chequered fashion. In the same way, we define parity of state $(i, j)$ that can be easily computed as the parity of coordinate sum: $(i + j) \mod 2$. Due to this parity of the bipartite graph, associated probabilities also depicts full transfusion between parts. However, the probability of staying inside of internal states decreases unevenly that leads to differences in the probability to being absorbed in even and odd states. Similarly, we analyzed the distinction of absorbing probabilities on even and odd turns.

The whole modelling analysis were applied to all three cases: BvB, PvE, PvP.

\subsection{Numerical simulations}

Although, the modelling and analytical analyses provides general information about statistics and probability properties of the game outcomes, individual trajectories remain hidden. To address this aspect we implement simulation of game process consisting in strategy picking from the arbitrary spatial distribution of mixed strategies and moving piece according to the rule matrix of possible moves. The simulation of piece movements starts at the initial state ($(0, 0)$ starting point). Next, at each turn the sampling of player choices is determined by the Eq.~(\ref{eq:sigma}) and carried out by Mersenne Twister pseudo-random number generator from \texttt{random} module of Python 3.8. Upon selecting the strategies, the piece movement is accomplished according to Eq.~(\ref{eq:transition}). Eventually, piece reaches the absorption state and the simulation stops. Finally, the simulation process produces trajectory of moves.

Further, we computed the similar statistical properties and distributions considered in the section "Game probability evolution modelling" to compare with the modelling and experimentally obtained statistics. All three approaches (numerical simulations, probability evolution modelling, inversion of fundamental matrix) and their statistics were implemented using Python 3.8 and performed on a work station (Core i5-8600 3.1GHz, 32Gb RAM). The source code is available at https://github.com/SermanVS/RWAnalyzer. 

\subsection{BvB analysis}
The game in case of BvB is a degenerate case and corresponds to the simple random walk on finite 2d lattice. However, we have not found the closed form of solution yet. Therefore, we applied the 3 described approaches (numerical simulations, probability evolution modelling, inversion of fundamental matrix) for the analysis of statistical properties of this case.

\subsection{PvE strategy analysis}
The case of playing with environment is nontrivial and implies the finding of optimal strategies for the player. The requirement of optimality presupposes the formulation of the problem in terms of global optimization in the space of mixed strategies. The objective function of the player is the mean absorption time ${\bf t_n^{PvE}}$ for the chosen mixed strategy $\sigma_{ij}^p$ defined in the Eq.~(\ref{eq:sigma}). That leads to the following two problem statements:
\begin{equation}
    \begin{aligned}
    {\bf t_n^{PvEA}} = \max_{f^A \in {\bf F}} {\bf t_n^{PvE}}, \hspace{1em} f_{ij}^B=1/2;
    \label{eq:centeropt}
    \end{aligned}
\end{equation}
\begin{equation}
    \begin{aligned}
    {\bf t_n^{PvEB}} = \min_{f^B \in {\bf F}} {\bf t_n^{PvE}}, \hspace{1em}  f_{ij}^A=1/2;
    \label{eq:borderopt}
    \end{aligned}
\end{equation}

where ${\bf F}$ is a space of vectors with elements in the range $[0, 1]$ that corresponds to the chosen relative frequencies $f_{ij}^p$ in the mixed strategy. To compute objective function ${\bf t_n^{PvE}}$ from certain frequencies we apply the Markov Chain framework (Eq.~(\ref{eq:P}), Eq.~(\ref{eq:N}), Eq.~(\ref{eq:T})). 

Small dimension problems can be solved directly by finding global optimum. To do this we apply \texttt{Minimize} function in the Wolfram Mathematica to lattice with size $5, 7$. In case of $3 \times 3$ field the time of absorption equal to 1 regardless of picked strategy. The case of $5 \times 5$ can be solved directly by \texttt{Minimize} function on the 9 parameters.
However, the large field sizes requires a lot of computation. To reduce the complexity we take advantage of symmetries in the game with respect to the main and side diagonals. According to this field collapses into upper triangle between diagonals of square lattice. The upper border of this triangle as previous is consist of absorption states. Other two sides of triangle belonging to diagonals provides reflective borders, i.e. piece movement on the diagonals continues inside the triangle. Such a reduction decreases the number of optimized parameters by 4 times. 

Though, the optimization problem were solved for smaller sizes, the deep investigation requires development of theoretical framework. Based on smaller results we suggests hypothesis that determines specific functional relation between field sizes and mean absorption times. As the first insight we described the algorithm of constructing mixed strategies that reaches the expected optimal times.


\section{Results}

Aiming at analysis process, Python 3.8 is used for modelling, simulations and inversion matrix computing. Results are addressed the 3 cases of the random walk game of 2 players on the finite lattice. Comparison to experimentally obtained trajectories of real games were done with using the $17 times 17$ field. The size of the field was chosen due to the comfortable game time: longer game leads to fatigue and smaller games has simpler strategies. The players were spent approximately 250 hours in total playing in the Random Walk Game to produce experimental data. In a result, we obtained trajectories of piece movements and corresponding player choices of 1562 real games in different modes (Cf. first row of Table~\ref{table:1}).

\subsection{Mean absorption time}
\begin{figure}[t]
    \begin{center}
    \includegraphics[width=\columnwidth,keepaspectratio,clip]{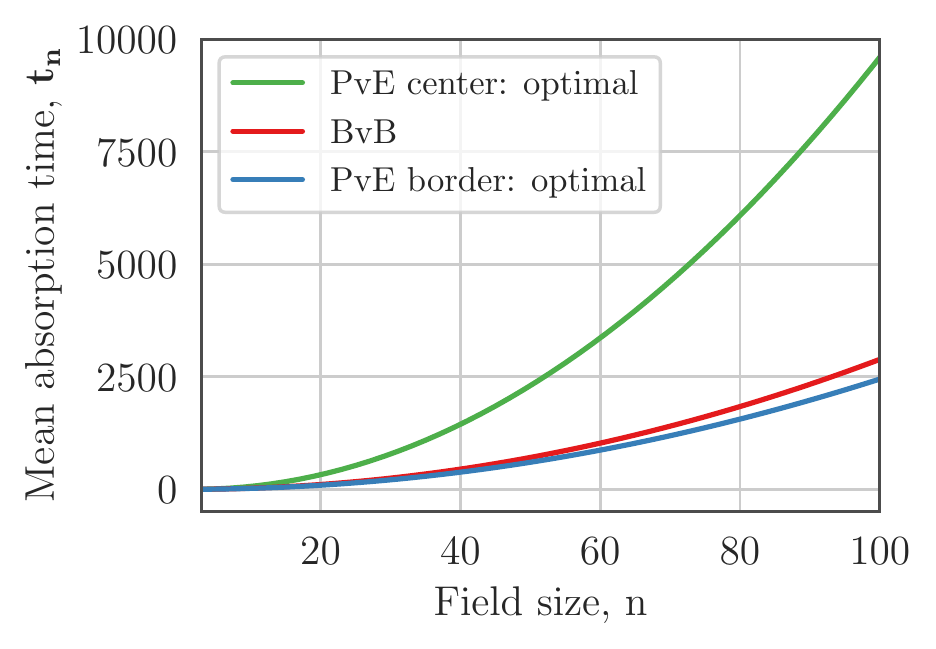}
    \caption{
        (Color online) Quadratic dependencies of mean absorption times on field sizes for cases: PvE center for player A (player has goal to stay inside as long as possible) -- green line, PvE border for player B (player has goal to reach border as soon as possible) -- blue line, and BvB (pure random walk absorption times) -- red line.
    }  
    \label{fig:2}
    \end{center}
\end{figure}

The analysis of BvB and PvE cases of random walk game suggests the quadratic function of the mean absorption time depending on the field size. The corresponding dependencies presented at Fig.~\ref{fig:2}. In case of PvE the identified relation for player A (goal: center) is ${\bf t_n^{PvE A}} = (n-2)^2$ and for player B (goal: border) is ${\bf t_n^{PvE B}} = ((n-1)/2)^2$. Though, relation for BvB were not found in closed form, it can be approximated by formula ${\bf t_n^{BvB}} = 0.294685413 n^2 - 0.232$ with mean absolute error less than $10^{-3}$ at the range $[3; 1001]$. Coefficients at leading term $n^2$ are approximately equal to $1.0$, $0.3$, $0.25$ in the order PvE goal center, PvE goal border, BvB. Asymptotically the expected time of the game in the optimal center strategy in PvE is 3 times longer than the absorption time in the pure random walk. However, the optimal border strategy in PvE is only about $20\%$ quicker than the pure random walk. 

Here, we showed the algorithm of the optimal strategy construction in PvE case. It is evident that increasing the number of states on one-dimensional segment leads to higher mean absorption times. Therefore, in case of requirement to reach the border as soon as possible it is profitable to choose the smallest one-dimensional segment on the 2d plane. There are 2 shortest segments that passes through the center: vertical and horizontal ({\color{red} Cf. Fig 3a }). According to the rules of movements (Cf. Fig 1), player B can explicitly choose one of those segments and keep the piece on this segment until the reaching border. The resulting process is a one-dimensional random walk on the shortest segment that has the property of shortest mean absorption time among other segments. The length of this segment is equal to $n$ and based on the random walk theory the mean absorption time is equal to ${\bf t_n^{PvE B}} = ((n-1)/2)^2$. Considering of other 2 dimensional strategies leads to increased number of states on the path to the absorption states resulting in increased mean absorption time. The formal proof were done only for small sizes $5, 7$ by solving the global optimization problem using Wolfram Mathematica. 

On the other hand, the goal of staying inside the field as long as possible requires the longest path for the random walk. The longest segment that can be placed within field corresponds to diagonal states. Although, such states are ambiguous, their constitute main and side diagonal "stairs" ({\color{red} Cf. Fig 3b}). The movement rules enable player A to keep piece only on the main diagonal regardless of choices of the second player. That's produces a random walk on the diagonal with length $2n-3$ because the states under the main diagonal are have symmetry according to movement rules. Then, mean absorption time is equal to ${\bf t_n^{PvE A}} = ((2n-3-1)/2)^2 = (n-2)^2$.

Another strategy for player A can be described as follows: in neighbors of absorption states choose pure strategy with one probability corresponding to movement away from the field border, and in other states choose any strategy. Analysis of this algorithm were done for the sizes $5, 7, 9, 11$. In a result of the $t_n$ simplification we found the independence of the mean absorption time from choices away from the border. The values of mean absorption times obtained by this algorithm matched to the optimal times of one-dimensional diagonal strategy ${\bf t_n^{PvE A}}$. The formal proof of optimality of this algorithm will be interesting part of the future works.

Next step in the analysis of the expected time of game endings is the comparison theoretical optimum to the experimental statistics. The detailed experimental study description were provided in the section "Experiment". In total we obtained 1062 games in the PvE mode: 528 of which are corresponds to playing with a center goal and 534 -- with a border goal. The participants who should have kept piece inside the field as long as possible in the average shows $145.45$ turns per one game. However, the optimal strategy suggests the mean absorption time equal to ${\bf t_{17}^{PvE A}} = (17-2)^2 = 225$, that $57\%$ higher than obtained in the real games. Although, players showed lower average times than optimal, their games are still 2 times longer than pure random walk that provides $75.2$ mean absorption time. The experiment shows the ability of the considered population to recognize properties of the game and to improve the simple strategy of random choices at each step regardless of knowledge about optimal strategies. However, the population strategy not converged to the optimal strategy in the average. Applying the modelling approach to the average population strategy in the PvE center mode provides similar result of $145.85$ absorption time.

Results of players aimed to reach border as soon as possible demonstrate the average number of turns equal to $71.12$. This is 7 turns ahead compared to the optimal strategy ${\bf t_n^{PvE B}} = ((17-1)/2)^2 = 64$ and 4 turns faster than pure random walk. Also, in that game mode we see slight improvement of equiprobable random choice strategy and gap with the optimal strategy. Eventually, modelling of the average population strategy for PvE border mode demonstrates quite higher mean absorption time $73.79$ in contrast to the modelling of PvE center strategy. This difference can be caused by inaccurate frequencies of rarely visited states.

Using frequencies from the PvE mode experimental games as input probabilities for the modelling process (Cf. section IV.B,C) and for the simulation procedure we observed correct reproducibility of results. Therefore, frequencies of corresponding strategy choices determined for each state allowed us to interpret set of $f_{ij}$ as the average strategy revealed by population. Simulation and modelling values are presented in the summary table of mean absorption times (Cf. Table~\ref{table:1}).

\begin{table}[]
    \setlength\extrarowheight{3pt}
    \begin{tabular}{|p{1.7cm}|C{1.2cm}|C{1.2cm}|C{1.2cm}|C{1.2cm}|C{1.2cm}|}
        \hline
        \multicolumn{1}{|c|}{\begin{tabular}[c]{@{}c@{}}Field size\\$17 \times 17$\end{tabular}} & \textbf{\begin{tabular}[c]{@{}c@{}}Random\\walk\end{tabular}} & \textbf{\begin{tabular}[c]{@{}c@{}}P vs B\end{tabular}} & \textbf{\begin{tabular}[c]{@{}c@{}}B vs P\end{tabular}} & \textbf{P vs P} & \textbf{\begin{tabular}[c]{@{}c@{}}P vs P\\ 400+\end{tabular}} \\ \hline
        \textbf{\# of games} & -     & 528    & 534   & 500    & 13     \\ \hline
        \specialrule{.1em}{.0em}{.0em}
        \textbf{Experiment}  & -     & 145.45 & 71.12 & 120.60 & 594.27 \\ \hline
        \textbf{Simulation}  & 75.22 & 145.77 & 73.66 & 115.93 & 132.91 \\
        \textbf{AMC theory}  & 75.21 & 145.85 & 73.79 & 116.22 & 133.22 \\
        \textbf{Modelling}   & 75.21 & 145.85 & 73.79 & 116.22 & 133.22 \\ \hline
        \textbf{Optimal}     & -     & 225.00 & 64.00 &    N/A & -      \\ \hline
    \end{tabular}
    \caption{
        The absorption mean times obtained by different approaches for the 3 game modes: BvB is a pure random walk on 2D finite lattice, PvE in case of two goals: keep piece inside the field as long as possible (center), reach the border as soon as possible (border), and PvP mode is a game of two players (all games and only games with length grater than 400 turns). The values are provided for the field size $17 \times 17$. The statistics in the simulation were performed using $10^5$ trajectories. The modelling were performed up to $10^4$ steps. The simulation, modelling and AMC theory (absorption Markov chain) were applied to the frequencies obtained from the real games in the experiment. In BvB case were used the equiprobable choice strategy.
    }
    \label{table:1}
\end{table}

Finally, results of participant games with each other (PvP) shows intermediate mean absorption times between PvE center and PvE border values. The average number of turns in PvP games is 57 moves longer than mean absorption time in the optimal strategy of the PvE mode with border goal. Comparison to the PvE mode with center goal shows 25 steps lower values in PvP versus experimental PvE games and approximately 2 times lower values in PvP versus optimal PvE center strategy.

Mean absorption times produced by modelling for particular frequencies are coincide with AMC theory mean values (error less than $10^{-9}$). Therefore, both methods can be applied to precisely compute mean absorption times for the arbitrary strategies. The simulation of $10^5$ trajectories shows good error, accurate to the integer part.

\subsection{Probability of evenness of the absorption time}
The modelling of the game process disclosed a difference between even and odd lengths of the game session. To analyse it we measure a probability to finish the game at even and odd turns.
Results are presented in Table~\ref{table:2}. Experimental results in the PvE with border goal depict almost equal frequencies of the ending at even and odd turns. However, in case of the center goal there are 2 times higher chances of ending up at the odd turn. The similar behavior were found in the PvP mode. These results indicate the bias to finish game on odd states (that sum of coordinates is odd). Although, odd ending states are observed more often, there is a approximately $30\%$ occurrences are ending at even states. This in turn suggests the necessity of nonzero probability of both odd and even endings in corresponding games between two players.

In the opposite to game of 2 players, the optimal strategies for PvE modes shows clear preferences to even endings in case of border goal and to odd endings in case of center goal. The border of the field can be firstly reached in 8 turns or $(n-1)/2$ for the arbitrary field size that corresponds to even numbers ($n$ is odd). Probability of such event for the optimal strategy is equal to $0.00815$ in the $17\times17$ field. The possibility to capture piece inside the field with one probability is available only for first $14$ turns and at $15$-th turn it will be absorbed with probability $\approx 0,000061$ in the $17\times17$ field. In general the first absorption turn corresponds to $n-2$ that is odd. 

\begin{table}[]
    \setlength\extrarowheight{3pt}
    \begin{tabular}{|p{1.7cm}|C{1.2cm}|C{1.2cm}|C{1.2cm}|C{1.2cm}|C{1.2cm}|}
        \hline
        \multicolumn{1}{|c|}{\begin{tabular}[c]{@{}c@{}}Field size\\$17 \times 17$\end{tabular}} & \textbf{\begin{tabular}[c]{@{}c@{}}Random\\walk\end{tabular}} & \textbf{\begin{tabular}[c]{@{}c@{}}P vs B\end{tabular}} & \textbf{\begin{tabular}[c]{@{}c@{}}B vs P\end{tabular}} & \textbf{P vs P} & \textbf{\begin{tabular}[c]{@{}c@{}}P vs P\\ 400+\end{tabular}} \\ \hline
        \textbf{Experiment} & -     & 30:70 & 51:49 & 35:65 & 15:84 \\ \hline
        \textbf{Simulation} & 50:50 & 30:70 & 51:49 & 36:64 & 31:69 \\
        \textbf{Modelling}  & 50:50 & 30:70 & 51:49 & 36:64 & 31:69 \\ \hline
        \textbf{Optimal}    & -     & 0:100 & 100:0 & N/A   & -     \\ \hline
    \end{tabular}
    \caption{
        The ratio between chances to finish the game at even number of turns relative to odd (even~$:$~odd). The values were obtained by different approaches for the 3 game modes: BvB is a pure random walk on 2D finite lattice, PvE in case of two goals: keep piece inside the field as long as possible (center), reach the border as soon as possible (border), and PvP mode is a game of two players. The values are provided for the field size $17 \times 17$. The statistics in the simulation were performed using $10^5$ trajectories. The modelling were performed up to $10^4$ steps. The simulation, modelling and AMC theory (absorption Markov chain) were applied to the frequencies obtained from the real games in the experiment. In BvB case were used the equiprobable choice strategy.
    }
    \label{table:2}
\end{table}

\subsection{Absorption time distributions}
\begin{figure*}[t]
    \begin{center}
    \includegraphics[width=\textwidth,keepaspectratio,clip]{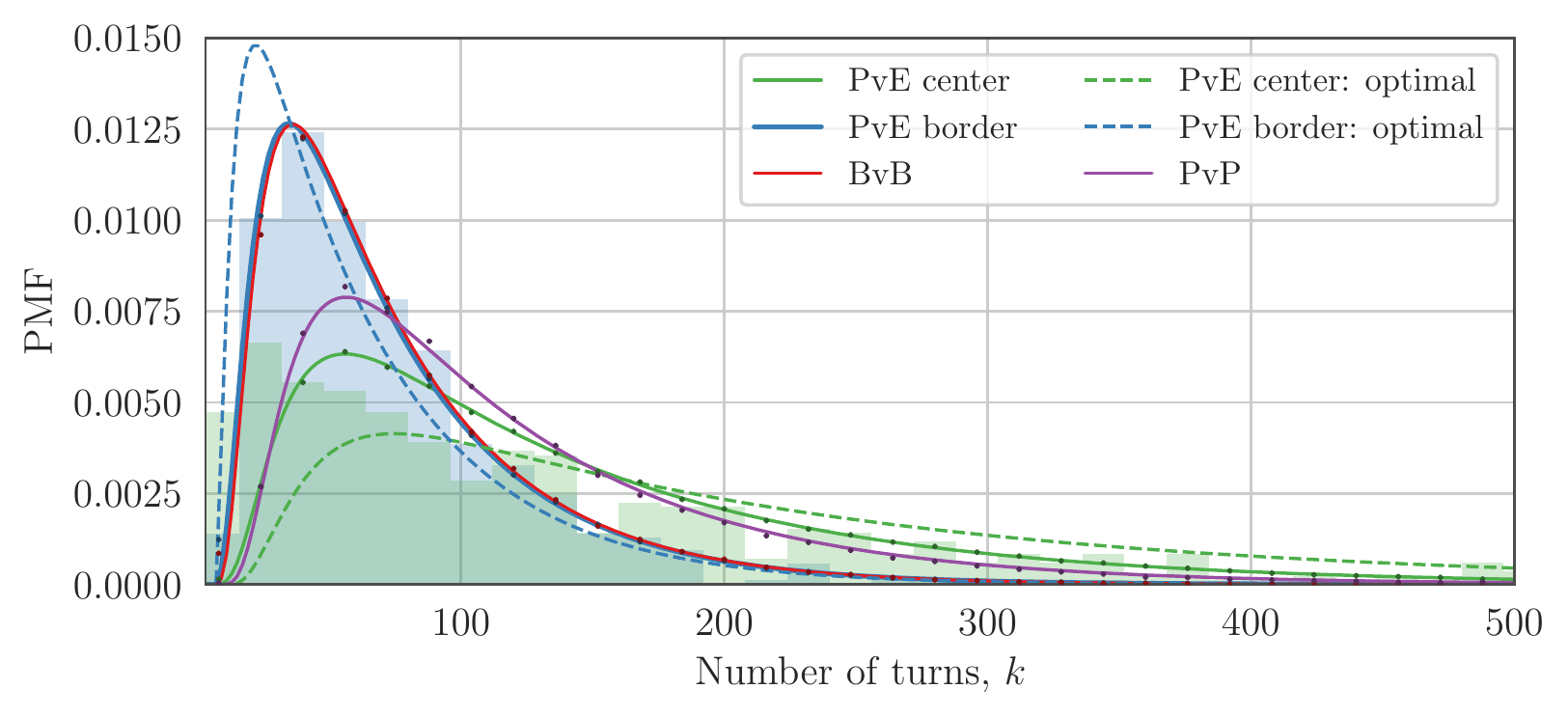}
    \caption{
        (Color online) Distributions of the number of turns obtained by modelling (solid line) and by simulation (dots) using corresponding strategies of players A and B. The BvB mode (red line) run equiprobable choice for both players. The PvE (green and blue) and PvP mode (purple line) curves were produced based on corresponding average population strategies. The opposite strategy in the PvE mode is the equiprobable choice strategy. The optimal strategy for the PvE center mode (green dashed line) is to keep a piece on the diagonal stair. The optimal strategy for the PvE border mode (blue dashed line) is to only chooses movements along the horizontal line. Experimentally obtained histograms are presented for PvE modes (green and blue area). 
    }  
    \label{fig:3}
    \end{center}
\end{figure*}

Both modelling and simulation approaches allowed us to compute precise and estimated probabilities of finishing the game at the arbitrary turn number. Applying of these methods require specifications of mixed strategy depends on the game field position. The pure random walk defines simple strategy of the equiprobable choice from pure strategies. Also, we analysed two suggested strategies of optimal random walk in the PvE mode. Besides of analytic strategies we collected experimental games of choice frequencies at each state. Based on those strategies we propagate probabilities through the field space and time. Additionally, we run simulations of $10^5$ individual trajectories using different strategies. In a result we obtain distributions of absorption times that presented on the Fig~\ref{fig:3}. 

The all distribution has differences in probabilities on even and odd turns on the close inspection. To assess only behavior associated with major trends, we consider histogram with wide bins (size 16). The all modes follows the same pattern of distribution: 
\begin{enumerate}
    \item the short games are rare;
    \item distribution has one statistic mode at an intermediate number of turns; 
    \item the probability of long games is exponentially decreases with game length increasing.
\end{enumerate}
The PvE center mode demonstrates similar distribution form except the statistic mode corresponds to short games that remain hidden with chosen bin size.

Although, the number of trajectories in the PvP mode are small (500), very long games were observed in the distribution with length greater than 400 turns. The probability of obtaining such games according to modelling the population average strategy is lower than $0.015$. However, 13 long trajectories were found in games of participants in the range from 461 to 964 turns and average absorption time equal to $594.27$. Discovered abnormality can be explained by "synchronization" between individual brain activities in long runs. As one turn takes 4.5 seconds in the average, game with 400 turns is 30 minutes long. Such a long time of the failure to finish the game by player B causes frustration and reduces concentration that can lead to unconscious decisions that can be easily predicted by player A. A loss of concentration results in worse ability of the individual to produce pure random independent choices. In this regard, such "synchronization" events can arise in the game.

\begin{figure*}[t]
    \begin{center}
    \includegraphics[width=\textwidth,keepaspectratio,clip]{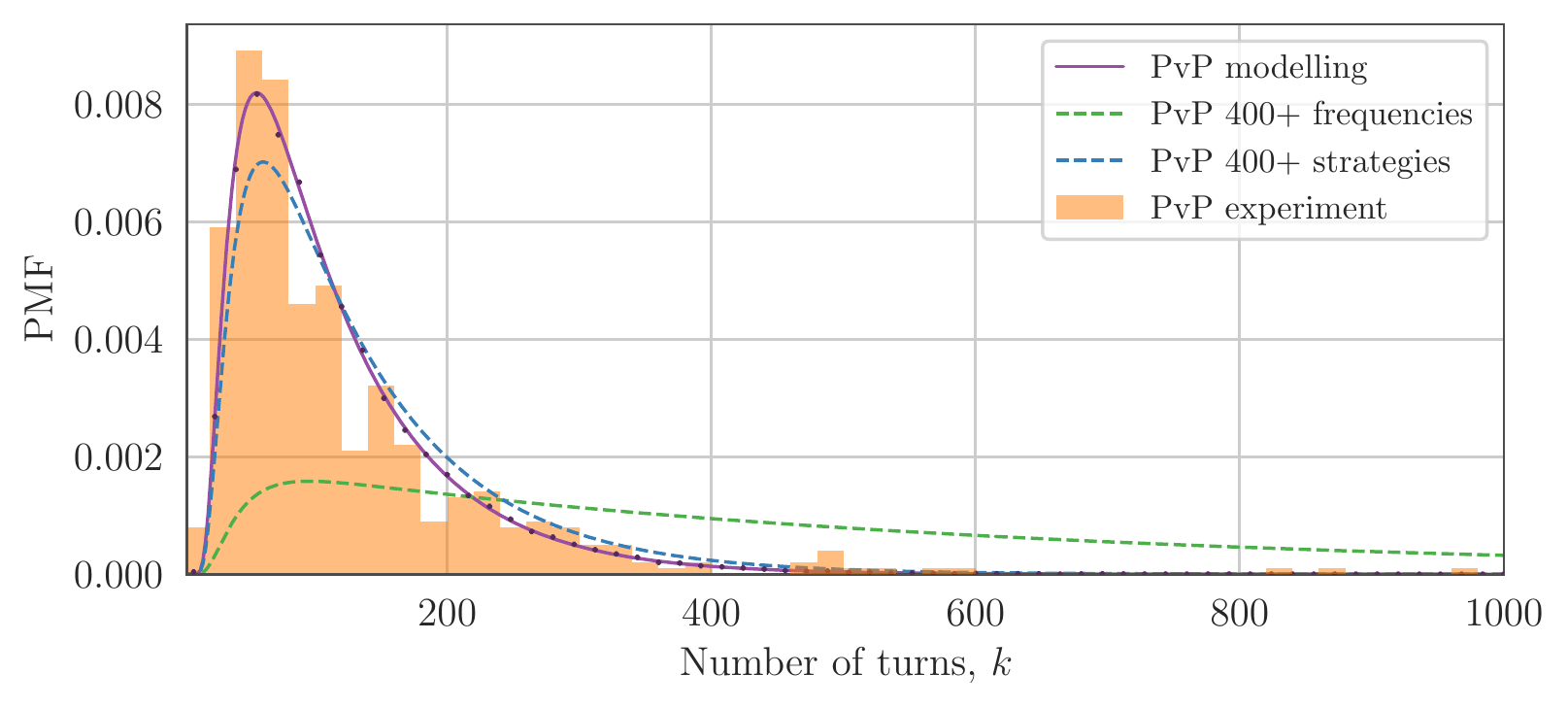}
    \caption{
        (Color online) The absorption time distributions for the PvP mode (yellow histogram and purple line) compared to modelling of movement frequencies (green line) and strategies (blue line) observed in long run games ($400+$ turns). The frequencies of movement directions for each state obtained in the experimental long run games were used to propagate probabilities over the field by the modelling approach. Strategies of both players A and B in the PvP with length over $400$ turns were used separately in the modelling approach.
    }  
    \label{fig:4}
    \end{center}
\end{figure*}

Deep analysis of those games was performed by decoupling strategies of those games from the main part of the distribution. Eventually, the modelling of the opposite average long run strategies produced the lower mean absorption time $\approx133.22$ turns. In order to compare distributions of such games we also modeled propagation of frequencies of movement directions that were observed in the long run games. The corresponding distributions of long run games are depicted in Fig~\ref{fig:4}. Although, the distribution of frequencies demonstrates the long tail, the underlying strategies that appears in the "synchronized" games did not reproduce the emergence of long games. Therefore, it demonstrates the presence of choice dependence on hidden factors that can not be explained only by the Markov chain property.

\subsection{Spatial distributions}
\begin{figure}[t]
    \begin{center}
    \includegraphics[width=\columnwidth,keepaspectratio,clip]{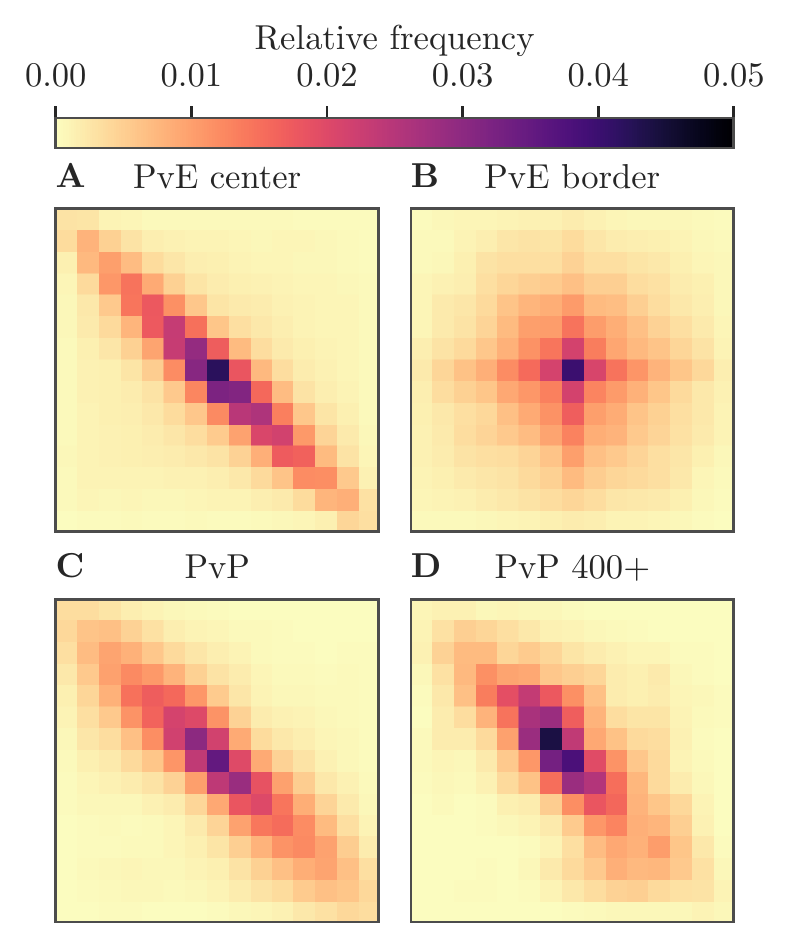}
    \caption{
        (Color online) The 2D distributions of state visits obtained in experimental games for 4 cases: PvE with center goal, PvE with the border goal, games of two players (PvP), and games longer than 400 turns in PvP mode.
    }  
    \label{fig:5}
    \end{center}
\end{figure}

Next, we analysed probabilities to find a piece in a particular state during the game. Similarly to previous sections we assessed experimentally obtained frequencies and results of modelling. Visualizations of spatial distributions for corresponding strategies are depicted on the Fig~\ref{fig:5}. 

The pure random walk demonstrates discrete Gaussian distribution over space. In contrast, addition of game rules changes resulting distribution picture. Predictably, the PvE mode with center goal demonstrates mostly movements on the diagonal (Cf. Fig~\ref{fig:5}A). Although, this behavior coincides with the spatial distribution in the suggested optimal strategy, the experimental data have higher spread around the diagonal states than optimal. 

The optimal strategy in PvE border mode should produce vertical or horizontal lines of involved states. In a result of experimental PvE border games we obtained 3 main patterns in the distribution (Cf. Fig~\ref{fig:5}B): the expected movements on the straight lines, and different from the optimal pattern is a distribution similar to the pure random walk on 2D square lattice. The second pattern shows the attempt of population to find the best strategy in the two-dimensional space rather than in one-dimensional line. However, the leaving of one-dimensional lines increases the mean absorption times. This explains the small difference in terms of mean absorption times between experimental PvE border and pure random walk (BvB). 

The spatial distribution pattern in the PvP mode (Cf. Fig~\ref{fig:5}C) is similar to the PvE center case: the piece mainly moves on the diagonal. The only difference is the enlarged spread around the diagonal line that shows stronger average strategy of player B with border goal compared to the equiprobable random choices (as in the PvE center mode). 

Last, we analysed the spatial distribution of states in games longer than 400 turns. In general, the probability of piece locations mostly concentrated on the main diagonal similar to previous cases (Cf. Fig~\ref{fig:5}D). Nevertheless, the distribution is more compact at the field center and has higher variation around the main diagonal compared to the PvE center and PvP cases. Such behavior suggests longer staying of piece close to the center during the long games, but also with movements from the diagonal and not only along it. 

\begin{figure*}[t]
    \begin{center}
    \includegraphics[width=\textwidth,keepaspectratio,clip]{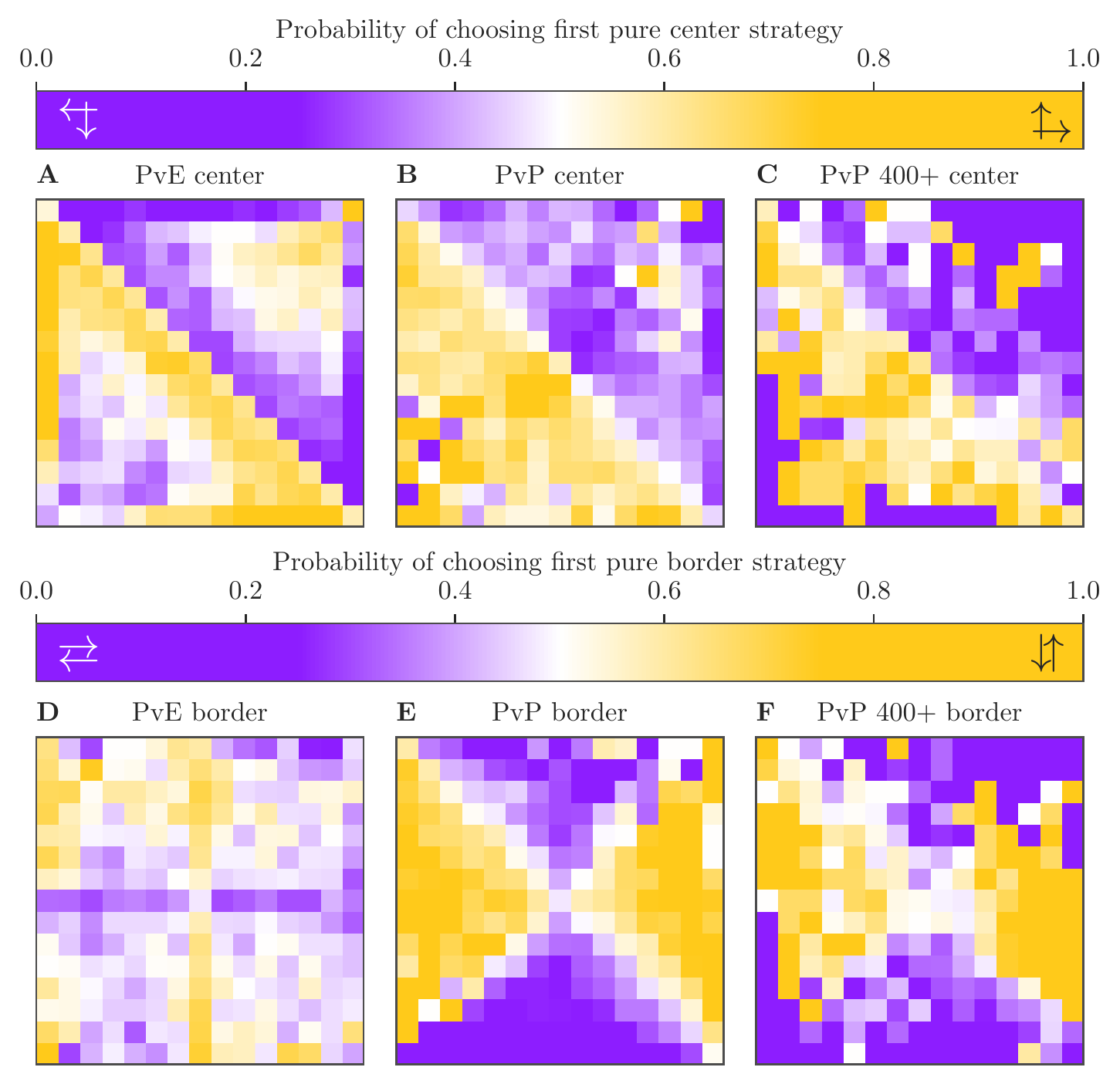}
    \caption{
        (Color online) The visualization of average population strategies for different modes obtained in the experiment. The color of cells depicts the frequency of choosing first pure strategy: for the center goal (A, B, C) and for the border goal (D, E, F). 
    }  
    \label{fig:6}
    \end{center}
\end{figure*}

\subsection{Strategy analysis}
Finally, we disentangle average population strategies of each player A and B and compare it to each other and to the optimal one. To present strategies we visualize them as the colored two-dimensional matrix with indices corresponding to the states on the 2D squared lattice (Fig.~\ref{fig:6}). Elements of the matrix depicts the frequency of choosing the first pure choice from possible choices according to the rules (Cf. Fig.~\ref{fig:1}). Technically, player A with center goal has two choices: move up/right, and move down/left; player B with border goal has another two choices: mode up/down, and move left/right. The divergent colors of elements demonstrate which pure strategy dominates at each state in the average over all games.

The observed strategy in case of PvE center mode shows mainly movement in direction of the diagonal states (Fig.~\ref{fig:6}A). However, states distant from the diagonal shows slightly higher frequencies of the opposite choices. Generally, players choose go away from the border, but in the states closer to corners on the side diagonal the behavior becomes more random (relative frequencies are closer to 0.5). The experimental strategy on the main diagonal looks similar to the first optimal strategy. In contrast, the choices on the border are differs from the second optimal strategy that suggests always moving away from the border. This results in a leak of probabilities outside the field not only in corners of main diagonal, but also in other border states.

The strategy of player with center goal in the PvP case also shows similar strategy to the PvE center mode (Fig.~\ref{fig:6}B). Moreover, players in the average choose to move in the direction of the main diagonal regardless of the position on the lattice. Almost all states at the border suggests similar strategy except some states with almost equal frequencies of both choices. In comparison to the PvE center mode, player in PvP mode with center goal have slightly lower confidence about moving towards main diagonal (the frequencies are closer to 0.5 in the PvP center compared to PvE center). 

Eventually, the PvE border strategy is well-defined at the horizontal and vertical lines (Fig.~\ref{fig:6}D). Although, players more often chose move towards the closest border at the central straight lines, frequencies in other states are not following to the common pattern. In contrast to the similarity of PvE and PvP center goal strategies, the strategy of PvP border demonstrates very different picture with clear pattern. In this case players act exactly the opposite to the PvE border mode. Their average strategy suggests the more often to choose moving along straight line that already conquered. That means movements along the coordinate line with lower deviation from the center. In a result plane of decisions divided into 4 alternated triangles. Although, the frequencies are close to 0 or 1, they are not equal to. The small difference demonstrates rare attempts of movements towards the closest border.

Last, strategies obtained in the long games between 2 players did not produce clear patterns due to limited number of games. However, there is similarity of decision principles to the normal PvP model \cite{3}. 

\begin{figure*}[t]
    \begin{center}
    \includegraphics[width=0.75\textwidth,keepaspectratio,clip]{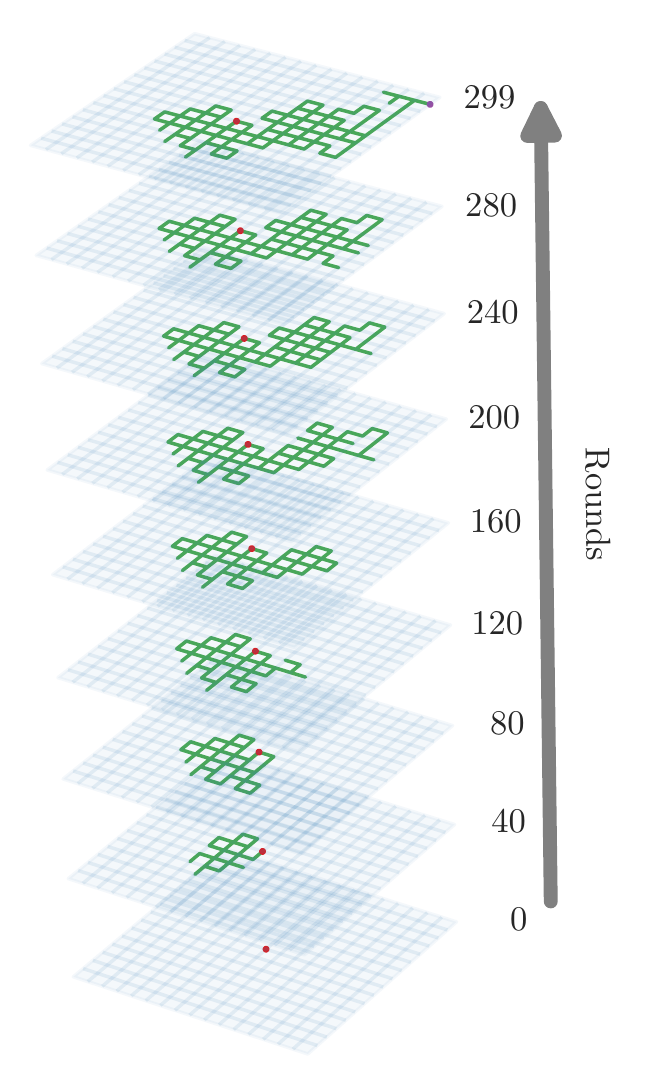}
    \caption{
        (Color online) Trajectory of PvP mode game obtained from the experiment divided by rounds that consist of 40 turns.
    }  
    \label{fig:7}
    \end{center}
\end{figure*}

\section{Conclusions}

We presented the results of experimental studies of a game-driven spatial process
which appears as the results of interaction of two antagonistic players. The collected trajectories were analyzed as the realizations of a two-dimensional random walk  in a finite domain.  We chose to focus on only on the statistics of survival time and their features such the shape of the corresponding probability distribution. 

We use a novel technique to conduct experiments and collect the data, based on a mobile app. Use of mobile technology to conduct research is a newest trend \cite{mobile}. We are not aware of any work on random walk processes or game theory where this technology has been used.

\begin{acknowledgments}
The authors acknowledge support of Russian Foundation for Basic Research No. 20-31-90121. 
\end{acknowledgments}

\section*{Data Availability}
The data reported in this paper are archived at Department of Applied Mathematics of Lobachevsky University and are available upon request.

\end{document}